\documentclass{article}
\usepackage{amsmath,amssymb, graphicx,url,times, enumitem, setspace}
\usepackage{spconf,amsmath,graphicx}

\usepackage{color}
\usepackage{multirow,bigdelim}
\usepackage{bm}
\usepackage{subfigure}
\usepackage{nccmath}

\usepackage{amsmath}
\usepackage{hyperref}
\input{def.set}
\usepackage{caption} 
\usepackage{algorithmic, algorithm}

\usepackage{fancyhdr}
\fancypagestyle{firstpage}{%
\pagestyle{fancy}
  \fancyhf{}
  
\rhead{\footnotesize{}}
\lhead{\footnotesize{Accepted in Proceedings of the ICASSP, Barcelona, Spain, May 4-8, 2020.}}
}

\title{Hybrid end-to-end Network with Context Aggregation  For Supervised \\Monaural Speech Enhancement}

\title{Dual Level Context Aggregation with Hybrid End-To-End Networks  For  \\ Supervised Monaural Speech Enhancement}

\title{Two-Staged Context Aggregation with Hybrid End-To-End Networks  For  \\ Supervised Monaural Speech Enhancement}

\title{Utterance based Temporal Context Aggregation in Two Stages
For Supervised \\ Monaural Speech Enhancement}

\title{ Temporal Context Aggregation in Two Stages
For Utterance based Monaural Speech Enhancement}

\title{Efficient Temporal Context Aggregation Using a Fully Convolutional and Recurrent Network for End-to-End Speech Enhancement}

\title{Efficient Context Aggregation for End-to-End Speech Enhancement\\Using a Densely Connected Convolutional and Recurrent Network}

\title{ A Dual-staged Context Aggregation method \\ Towards Efficient End-to-End Speech Enhancement}

\name{Kai Zhen$^{1,2}$, Mi Suk Lee$^3$, Minje Kim$^{1,2}$\thanks{This work was supported by Institute for Information \& communications Technology Promotion (IITP) grant funded by the Korea government (MSIT) (2017-0-00072, Development of Audio/Video Coding and Light Field Media Fundamental Technologies for Ultra Realistic Tera-media).}}
\address{$^1$Indiana University, Luddy School of Informatics, Computing, and Engineering, Bloomington, IN\\
$^2$Indiana University, Cognitive Science Program, Bloomington, IN\\
$^3$Electronics and Telecommunications Research Institute, Daejeon, South Korea\\
\texttt{zhenk@indiana.edu},~ \texttt{lms@etri.re.kr},~ \texttt{minje@indiana.edu}}

\begin{document}

\ninept
\maketitle
\thispagestyle{firstpage}

\begin{sloppy}

\begin{abstract}

In speech enhancement, an end-to-end deep neural network converts a noisy speech signal to a clean speech directly in the time domain without time-frequency transformation or mask estimation. However, aggregating  contextual information from a high-resolution time domain signal with an affordable model complexity still remains challenging. 
In this paper, we propose a densely connected convolutional and recurrent network (DCCRN), a hybrid architecture, to enable dual-staged temporal context aggregation. 
With the dense connectivity and cross-component identical shortcut, DCCRN consistently outperforms competing convolutional baselines with an average STOI improvement of 0.23 and PESQ of 1.38 at three SNR levels. The proposed method is computationally efficient with only 1.38 million parameters. The generalizability performance on the unseen noise types is still decent considering its low complexity, although it is relatively weaker comparing to Wave-U-Net with 7.25 times more parameters.

\end{abstract}

\begin{keywords}
End-to-end, speech enhancement, context aggregation, residual learning, dilated convolution, recurrent network
\end{keywords}

\section{Introduction}
\label{sec:intro}


Monaural speech enhancement can be described as a process to extract the target speech signal by suppressing the background interference in the speech mixture in the single-microphone setting. 
There have been various classic methods, such as spectral subtraction \cite{BollSF79ieeeassp}, Wiener-filtering \cite{brown1992introduction} and non-negative matrix factorization \cite{schmidt2006single}, to remove the noise without leading to objectionable distortion or adding too much artifacts, such that the denoised speech is of decent quality and intelligibility. Recently, the deep neural network (DNN), a data-driven computational paradigm, has been extensively studied thanks to its powerful parameter estimation capacity and correspondingly promising performance \cite{williamson2016complex}\cite{kim2019incremental}\cite{luo2019conv}\cite{xu2014regression}.

DNNs formulate monaural speech enhancement either as mask estimation \cite{narayanan2013ideal} or end-to-end mapping \cite{pascual2017segan}. In terms of mask estimation, DNNs usually take acoustic features in time-frequency (T-F) domain to estimate a T-F mask, such as  ideal binary mask (IBM) \cite{wang2005ideal}, etc. In comparison, both the input and output of end-to-end speech enhancement DNNs can be T-F spectrograms, or even time domain signals directly without any feature engineering. 

In both mask estimation and end-to-end mapping DNNs, dilated convolution \cite{yu2015multi} serves a critical role to aggregate contextual information with the enlarged receptive field. Gated residual network (GRN) \cite{tan2018gated} employs dilated convolutions to accumulate context in temporal and frequency domains, leading to a better performance than a long short-term memory (LSTM) cell-based model \cite{chen2017long}. In end-to-end setting, WaveNet \cite{oord2016wavenet} and its variations also adopt dilated convolution in speech enhancement.


For real-time systems deployed in resource-constrained environment, however, the oversized receptive field from dilated convolution can cause a severe delay issue. Although causal convolution can enable real-time speech denoising \cite{tan2018convolutional}, it performs less well comparing to the dilated counterpart \cite{tan2018gated}. Besides, when the receptive field is too large, the amount of padded zeroes in the beginning of the sequence and a large buffer size for online processing can be a burdensome spatial complexity for a small device. Meanwhile, recurrent neural networks (RNN) can also aggregate context through
a frame-by-frame processing without relying on the large receptive field. However, the responsiveness of a practical RNN system, such as LSTM \cite{chen2017long}, comes at the cost of the increased number of model parameters, which is neither as easy to train nor resource-efficient. 
There has been effort to apply dilated DenseNet \cite{li2019densely} or a hybrid architecture to source separation \cite{takahashi2018mmdenselstm}, the mechanism to enable dual-staged context aggregate through the heterogeneous model topology has not been addressed.



%

To achieve efficient end-to-end monaural speech enhancement, we propose a densely connected convolutional and recurrent network (DCCRN), which conducts dual-level context aggregation. 
The first level of context aggregation in DCCRN is achieved by a dilated 1D convolutional neural network (CNN) component, encapsulated in the DenseNet architecture \cite{huang2017densely}.
It is followed by a compact gated recurrent unit (GRU) component \cite{chung2014empirical} to further utilize the contextual information in the ``many-to-one" fashion. Note that we also employ a cross-component identical shortcut linking the output of DenseNet component to the output of GRU component to reduce the complexity of the GRU cells. We also propose a specifically designed training procedure for DCCRN that trains the CNN and RNN components separately, and then finetune the entire model. Experimental results show that the hybrid architecture of dilated DenseNet and GRU in DCCRN consistently outperforms other CNN variations with only one level of context aggregation on untrained speakers. Our model is computationally efficient and provides reasonable generalizability to untrained noises with only 1.38 million parameters. 


We describe the proposed method in Section \ref{sec:model}, and then provide experimental validation in Section \ref{sec:experiment}. We conclude in Section \ref{sec:conclusion}.



\begin{figure*}[t]
    \centering
    \includegraphics[width=\textwidth]{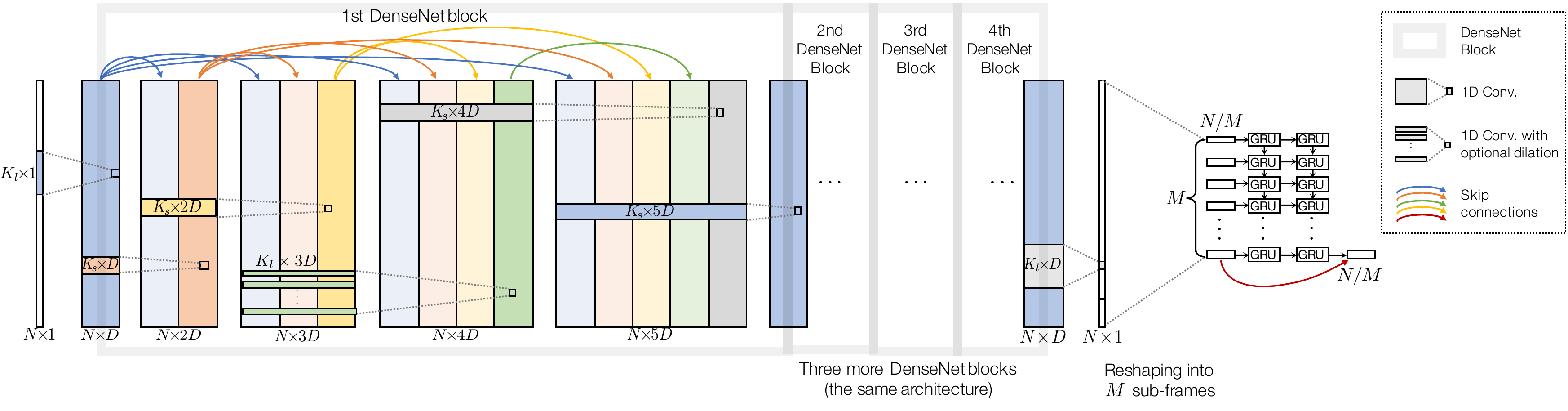}
    \caption{A schematic diagram of the DCCRN training procedure including dilated DenseNet and GRU components.}
    \label{fig:system}
\end{figure*}

\section{Model Description}
\label{sec:model}


\subsection{Context aggregation with dilated DenseNet}
Residual learning has become a critical technique to tackle the  gradient vanishing issue when tuning a deep convolutional neural network (CNN), such that the deep CNN can achieve better performance but with a lower model complexity. 
ResNet illustrates a classic way to enable residual learning by adding identical shortcuts across bottleneck structures \cite{he2016deep}. Although the bottleneck structure includes direct paths to feedforward information from earlier layers to later layers, it does not extend to its full capacity of the information flow. Therefore, ResNet is usually found to be accompanied by a gating mechanism, a technique heavily used in RNNs, such as LSTM or GRU, to further facilitate the gradient propagation in convolutional networks \cite{tan2018gated}. 




In comparison, DenseNet \cite{huang2017densely} resolves the issue by redefining the skip connections. The dense block differs from the bottleneck structure in that each layer takes concatenated outputs from all preceding layers as its input, while its own output is fed to all subsequent layers (Figure \ref{fig:system}). Consequently, DenseNet requires fewer model parameters to achieve a competitive performance. 

In fully convolutional architectures, the dilated convolution is a popular technique to enlarge the receptive field to cover longer sequences \cite{oord2016wavenet}, which has shown promising results in speech enhancement \cite{tan2018gated}. Because of the lower model complexity, dilated convolution is considered as a cheaper alternative to the recurrence operation. Our model adapts this technique sparingly with a receptive field size that does not exceed the frame size.  


We use $\calF^{(l)}$ to denote a convolution operation between the input $\bX^{(l)}$ and the filter $\bH^{(l)}$ in the $l$-th layer with a dilation rate $\gamma^{(l)}$:
\begin{align}
\label{eq:lthconv}\bX^{(l+1)}&\leftarrow\calF^{(l)}(\bX^{(l)},  \bH^{(l)}, \gamma^{(l)})\\
\label{eq:dilatedConv}\bX^{(l+1)}_{\tau, d}&=\sum_{n+k\gamma^{(l)} =\tau}\bX_{n, d}^{(l)}\bH_{k, d}^{(l)}, 
\end{align}
where $n, \tau, d, k$ are the indices of the input features, output features, channels, and filter coefficients, respectively. Note that $k$ is an integer with a range $k \le\lfloor K/2 \rfloor$, where $K$ is the 1D kernel size. In our system we have two kernel sizes: $K_s=5$ and $K_l=55$. As DCCRN is based on 1D convolution, the tensors are always in the shape of (features)$\times$(channels). Zero padding keeps the number of features the same across layers. Given the dilation rate $\gamma^{(l)} > 1$ \cite{yu2015multi}, the convolution operation is defined in \eqref{eq:dilatedConv} with the dilation being activated. 

In DCCRN, a dense block combines five such convolutional layers. In each block, the input to the $l$-th layer is a channel-wise concatenated tensor of all preceding feature maps in the same block, thus substituting \eqref{eq:lthconv} with
\begin{align}
\label{eq:denseconv}\bX^{(l+1)}&\leftarrow\calF^{(l)}\Big(\big[\bX^{(l)}, \bX^{(l-1)},\cdots, \bX^{(l_b)}\big] ,  \bH^{(l)}, \gamma^{(l)}\Big),
\end{align}
where $\bX^{(l_b)}$ denotes the first input feature map to the $b$-th block. Note that in this DenseNet architecture, $\bH^{(l)}$ grows its depth accordingly, i.e., $\bH^{(l)}\in \Real^{K\times (l-l_b+1)D }$ with a growing rate $D$, the depth of $\bX^{(l_b)}$. In the final layer of a block, the concatenated input channels collapse down to $D$, which forms the input to the next block. The first dense block in Figure \ref{fig:system} depicts this process. We stack four dense blocks with the dilation rate of the middle layer in each block to be 1, 2, 4 and 8, respectively. Different from the original DenseNet architecture, we do not apply any transition in-between blocks, except for the very first layer, prior to the stacked dense blocks, expanding the channel of the input from $1$ to $D$, and another layer right after the stacked dense blocks to reduce it back to $1$. This forms our fully convolutional DenseNet baseline. In all the convolutional layers, we use leaky ReLU as the activation.







\subsection{Context aggregation with gated recurrent network}

DCCRN further employs RNN layers following the dilated DenseNet component (Figure \ref{fig:system}). Among LSTM and GRU, two most well-known RNN variations, DCCRN chooses GRU for its reduced computational complexity compared to LSTM. The information flow within each unit is outlined as follows:
\begin{align}
   \label{eq:update}\bh(t) &= \big(1-\bz(t)\big)\odot \bh(t-1)+\bz(t) \odot \tilde{\bh}(t)\\
   \label{eq:hidden}\tilde{\bh}(t) &= \tanh\Big(\bW_h\bx(t) + \bU_h\big(\br(t)\odot \bh(t-1)\big)\Big) \\
   \label{eq:updateG} \bz(t) &= \sigma\big(\bW_z\bx(t) + \bU_z\bh(t-1)\big)\\
    \label{eq:resetG}\br(t) &= \sigma\big(\bW_r\bx(t) + \bU_r \bh(t-1)\big),
\end{align}
where $t$ is the index in the sequence. $\bh$ and $\tilde{\bh}$ are the hidden state and the newly proposed one, which are mixed up by the update gate $\bz$ in a complementary fashion as in \eqref{eq:update}. The GRU cell computes the tanh unit $\tilde{\bh}$ by using a linear combination of the input $\bx$ and the gated previous hidden state $\bh$  as in \eqref{eq:hidden}. Similarly, the gates are estimated using another sigmoid units as in \eqref{eq:updateG} and \eqref{eq:resetG}. In all linear operations, GRU uses corresponding weight matrices, $\bW_h, \bW_z, \bW_r, \bU_h, \bU_z, \bU_r$. We omit bias terms in the equations. 

The GRU component in this work follows a ``many-to-one" mapping style for an additional level of context aggregation. During training, it looks back $M$ time steps and generates the output corresponding to the last time step. To this end, DCCRN reshapes the output of the CNN part, the $N\times 1$ vector, into $M$ sub-frames, each of which is an $N/M$-dimensional input vector to the GRU cell. We have two GRU layers, one with 32 hidden units and the other one with $N/M$ units to match the output dimensionality of the system. Furthermore, to ease the optimization and to limit the model complexity of the GRU layers, we pass the last $N/M$ sub-frame output of the DenseNet component to the output of GRU component via a skipping connection, which is additive as in the ResNet architecture---the denoised speech is the sum of the output from both components. With the dilated DenseNet component well-tuned, its output will already be close to the clean speech, which leaves less work for GRU to optimize, as detailed in Section \ref{sec:training}.


\subsection{Data flow}

During training, as illustrated in Figure \ref{fig:system}, the noisy frame is first fed to the DenseNet component $\calD(\bx; \mathbb{W}^{\text{CNN}})$ (line \ref{alg:cnn} in Algorithm \ref{algo:1}). It comprises of $L$ consecutive convolutional layers that are grouped into four dense blocks, where $\mathbb{W}^{\text{CNN}}=\{\bW^{(1)}, \cdots, \bW^{(L)}\}$. The output frame of DenseNet, containing $N$ samples, is then reformulated to a sequence of $N/M$ dimensional vectors, $\bar{\bX}^{(L)}\in\Real^{N/M\times M}$, which serve as the input of the GRU component: $\hat{\bs} \leftarrow \calG(\bar{\bX}^{(L)}; \mathbb{W}^{\text{RNN}})$. The cleaned-up signal $\hat{\bs}$  corresponds to the final state of the GRU  with the dimension of  $N/M$. 

At test time, the output sub-frame of DCCRN is weighted by Hann window with $50\%$ overlap by its adjacent sub-frames. Note that to generate the last $N/M$ samples, DCCRN only relies on the current and past samples, up to $N$ within that frame, without seeing future samples, which is similar to causal convolution. Hence, the delay of DCCRN is the sub-frame size ($256/16,000=0.016$ second). If it were just for the DenseNet component only, such as those convolutional baselines compared in Section \ref{sec:experiment}, the Hann window with the same overlap rate would still be applied, but the model output would be all $N$ samples for the corresponding frame, instead of the last $N/M$ samples.

Table \ref{tab:topo} summarizes the network architecture. The current topology is designed for speech sampled at 16kHz.

\begin{algorithm}[t]
\caption{The feedforward procedure in DCCRN}
\label{algo:1}
\begin{algorithmic}[1]
\STATE \textbf{Input:} $N$ samples from the noisy utterance, $\bx$\\
\STATE \textbf{Output:} The last $M/N$ samples of the denoised signal, $\hat{\bs}$\\
\STATE DenseNet denoising: $\bX^{(L)} \leftarrow \calD(\bx; \mathbb{W}^{\text{CNN}})$\\\label{alg:cnn}
\STATE Reshaping: $\bar{\bX}^{(L)} \leftarrow \big[\bX^{(L)}_{1:N/M}, \bX^{(L)}_{N/M+1:2N/M}, \cdots,$\\
\nonumber \STATE \hspace{1.2in} $\bX^{(L)}_{N-N/M+1:N}\big] $\\
  \STATE GRU denoising: $\hat{\bs} \leftarrow \calG(\bar{\bX}^{(L)}; \mathbb{W}^{\text{RNN}})$\\
  \STATE Post windowing: $\hat{\bs} \leftarrow \text{Hann}(\hat{\bs})$\hspace{0.1in}\COMMENT{\# at test time only}\\
 \end{algorithmic}
\end{algorithm}

\begin{table}[t]
\centering
\caption{Architecture of DCCRN: for the CNN layers, the data tensor sizes are represented by (size in samples, channels), while the CNN kernel shape is (size in samples, input channels, output channels). For the GRU layers, an additional dimension for the data tensors defines the length of the sequence, $M=4$, while the kernel sizes define the linear operations (input features, output features). The middle layer of each dense block, marked by a dagger, is with larger kernel size $K_l=55$ and an optional dilation with the rate of 1, 2, 4, and 8, for the four dense blocks, respectively.}
\setlength\tabcolsep{3.0pt}
\begin{tabular}{ c|c|c|c }
 \hline
 Components &Input shape & Kernel shape & Output shape\\
 \hline
Change channel & (1024, 1) & (55, 1, 32) &(1024, 32) \\
\hline
DenseNet & (1024, 32) & \begin{tabular}{cc}\rule[6pt]{0pt}{0pt}(5, 32, 32)  & \rdelim]{5}{5mm}[$\times$4]\\ (5, 64, 32) \\ (55, 96, 32)$^{\dagger}$  &  \\(5, 128, 32)\\ (5, 160, 32) &\rule[-2pt]{0pt}{0pt}\end{tabular} &(1024, 32) \\\hline
Change channel & (1024, 32) & (55, 32, 1) &(1024, 1) \\\hline
Reshape & (1024, 1) & - &(4, 256, 1) \\\hline
GRU & (4, 256, 1) & \begin{tabular}{c}(256+32, 32)$\times 3$ \\ (32+256, 256)$\times 3$ \end{tabular}  &(256, 1) \\\hline
\end{tabular}
\vspace{-0.05in}
\label{tab:topo}
\end{table}

\subsection{Objective function}


It is known that the mean squared error (MSE) itself cannot directly measure the perceptual quality nor
the intelligibility, both of which are usually the actual metrics for evaluation. To address the discrepancy, the
MSE can be replaced by a more intelligibly salient
measure, such as short-time objective intelligibility
(STOI) \cite{fu2018end}. However, the improved
intelligibility does not guarantee a better perceptual
quality. The objective function in this work is defined in \eqref{eq:obj}, which is still
based on MSE, but accompanied by a regularizer that
compares mel spectra between the target and output signals. The TF domain regularizer compensates the end-to-end DNN that would only operate in time domain, otherwise.  Empirically, it is shown to achieve better perceptual quality, as proposed in \cite{zhen2019cascaded}.

\begin{align}
    \calE(\bs||\hat{\bs}) = \text{MSE}(\bs||\hat{\bs}) + \lambda \text{MSE}\big(\text{Mel}(\bs)||\text{Mel}(\hat{\bs})\big).
    \label{eq:obj}
\end{align}

\subsection{Model training scheme}\label{sec:training}
We train the CNN and RNN components separately, and then finetune the combined network.\vspace{-0.05in}
\begin{itemize}[leftmargin=0in]\setlength{\itemindent}{.15in}
\item {\em\textbf{CNN training:}} First, we train the CNN component to minimize the error $\argmin\limits_{\mathbb{W}^{\text{CNN}}} \calE(\by||\calD(\bx; \mathbb{W}^{\text{CNN}}))$. \vspace{-0.05in}

\item {\em\textbf{RNN training:}} Next, we train the RNN part by minimizing $\argmin\limits_{\mathbb{W}^{\text{RNN}}} \calE(\bs||\calG(\bar{\bX}^{(L)}; \mathbb{W}^{\text{RNN}}))$, while $\mathbb{W}^{\text{CNN}}$ is locked. \vspace{-0.05in}

\item {\em\textbf{Integrative finetuning:}} Once both CNN and RNN components are pretrained, we unlock the CNN parameter and finetune both components to minimize the final error: $\argmin\limits_{\mathbb{W}^{\text{CNN}}, \mathbb{W}^{\text{RNN}}} \calE(\bs||\calG\Big(\calD\big(\bx; \mathbb{W}^{\text{CNN}}\big); \mathbb{W}^{\text{RNN}}\Big)$.
Note that the learning rate for integrative finetuning should be smaller. \vspace{-0.05in}

\end{itemize}

\begin{table*}[t]
\caption{SDR, SIR, SAR, STOI and PESQ comparison on untrained speakers and trained noise types}
\vspace{-0.1in}
\centering
\resizebox{\textwidth}{!}{
\begin{tabular}{ |c|c|c|c|c|c|c|c|c|c|c|c|c|c|c|c|c|c|c|c|c| }
 \hline
  {Metrics} &\multicolumn{3}{c|}{{SDR (dB)}}& \multicolumn{3}{c|}{{SIR (dB)} }& \multicolumn{3}{c|}{{SAR (dB)} }& \multicolumn{3}{c|}{{STOI (\%)} }& \multicolumn{3}{c|}{{PESQ} }\\
  \hline
  {SNR level (dB)} &{-5}&{0}&{5}&{-5}&{0}&{5}&{-5}&{0}&{5}&{-5}&{0}&{5}&{-5}&{0}&{5}\\
\hline
{Unprocessed}  &-5.01 & -0.00 & 5.01&
-4.97 & 0.04 & 5.06&
25.43 & 27.97 & 30.56&
0.65 & 0.72 & 0.78&
1.06 & 1.11 & 1.21\\
\hline
{Dilated DenseNet}     &13.54 & 15.67 & 17.35&
19.78 & 21.15 & 22.77&
14.85 & 17.25 & 19.22&
0.88 & 0.92 & 0.94&
1.95 & 2.32 & 2.58\\
{DenseNet+GRU}&13.89 & 16.63 & 18.72&
20.92 & 23.21 & 25.28&
14.78 & 17.17 & 19.68&
0.90 & 0.93 & 0.95&
1.96 & 2.35 & 2.57\\
{DCCRN}  &\textbf{15.11}&\textbf{17.51} &{19.29}
&\textbf{21.42} &{24.64} &\textbf{26.70}
&{16.08} &\textbf{18.82} &{20.77}
&\textbf{0.92}&\textbf{0.95}&\textbf{0.96}
&\textbf{2.14}&\textbf{2.55}&\textbf{2.83}\\

{DCCRN*}&{15.08} & {17.45} & \textbf{19.31}&
{21.30} & \textbf{24.71} & {26.33}&
\textbf{16.22} & {18.76} & \textbf{20.83}&
\textbf{0.92}  & {0.94}  & \textbf{0.96}&
{2.13}  & {2.54}  & {2.82}
\\\hline
\end{tabular}}
\label{tab:sss}
\end{table*}

\section{Experiments}
\label{sec:experiment}
\subsection{Experimental setup}
In this paper, the experiment runs on TIMIT corpus \cite{timit}.
We consider two experimental settings.
For the model training, we randomly select 1000 utterances from TIMIT training subset. 5 types of non-stationary noise (birds, cicadas, computer keyboard, machine guns and motorcycles) from \cite{vincent2006performance} are used to create mixtures. Concretely, each clean signal is mixed with a random cut of each of these noise types at a SNR level randomly drawn from the set of integers with the range of $[-5, +5]$ dB.
Therefore, 5,000 noisy speech samples, totaling 3.5 hours, are used for model training.
At test time, we randomly select 100 unseen utterances from TIMIT test subset, and mix each utterance with those 5 types of noise to construct a test set of with unseen speakers. 
The noise is randomly cut to match the length of test utterances. The mixtures are generated from 3 SNR levels (-5 dB, 0 dB and +5 dB), yielding 1500 test utterances in total.

\subsection{Baselines}
To validate the two-staged context aggregation method, we compare DCCRN to regular DenseNets with one aggregation mechanism. 
\begin{itemize}[leftmargin=0in]\setlength{\itemindent}{.15in}
    \item \textbf{\em Dilated DenseNet} uses dilated convolution in the middle layer of each dense block, or DCCRN without context aggregation from GRU layers.\vspace{-0.04in}
    \item \textbf{\em DenseNet+GRU} refers to the DenseNet architecture coupled with two GRU layers, but with no dilation. \vspace{-0.04in}
    \item \textbf{\em DCCRN} is our full model with the proposed training scheme in Section \ref{sec:training}. We train the CNN part with 100 epochs, the GRU component with 20 epochs, followed by integrated finetuning of 20 epochs. The learning rates are 1e-4, 5e-6 and 5e-7, correspondingly. 
    \item \textbf{\em DCCRN}$^*$ is with an enlarged frame size, $N=4096$.
\end{itemize}

All models are trained to our best effort with Adam optimizer \cite{adam}.  The batch size is 32 frames. The regularizer coefficient, $\lambda$, is $1/60$. The GRU gradients are clipped in the range of $[-0.1, 0.1]$. \vspace{-0.05in}

\subsection{Performance analysis on untrained speakers}
To evaluate the performance, we use BSS\_Eval toolbox \cite{vincent2006performance}.
The BSS\_Eval toolbox provides an objective evaluation on source separation performance, by decomposing the overall error
signal-to-distortion ratio (SDR) into components of specific error types. In this work, we focus on signal to interference ratio (SIR), and signal to artifacts ratio (SAR).
We also choose short-time objective intelligibility (STOI) \cite{stoi} and perceptual evaluation of speech quality (PESQ) with the wide-band extension (P862.3) \cite{pesq} to measure the intelligibility and quality of the denoised speech. Note that narrow-band PESQ scores (P862) \cite{pesqnb} are approximately greater by 0.5 (e.g., $1.67$ for unprocessed utterances at 0 dB SNR in our case).


%
%
%
As shown in Table \ref{tab:sss}, by coupling both context aggregation techniques, DCCRN consistently outperforms the other baseline models in all metrics. In terms of SDR, the average improvement is $17.3\%$ comparing to the DenseNet baseline.
The comparison with unprocessed mixtures shows an average STOI improvement of $+0.23$ and PESQ of $+1.38$.
The performance is not further improved with $N=4096$, due to the trade-off between the increased difficulty in GRU optimization and more temporal context in each sequence\footnote{Denoised samples are available at 
\url{https://saige.sice.indiana.edu/research-projects/DCCRN}}.

\subsection{Generalizability for untrained speakers and noises}

\begin{figure}[h]
\centering
\subfigure[STOI]{\includegraphics[height=1.75in]{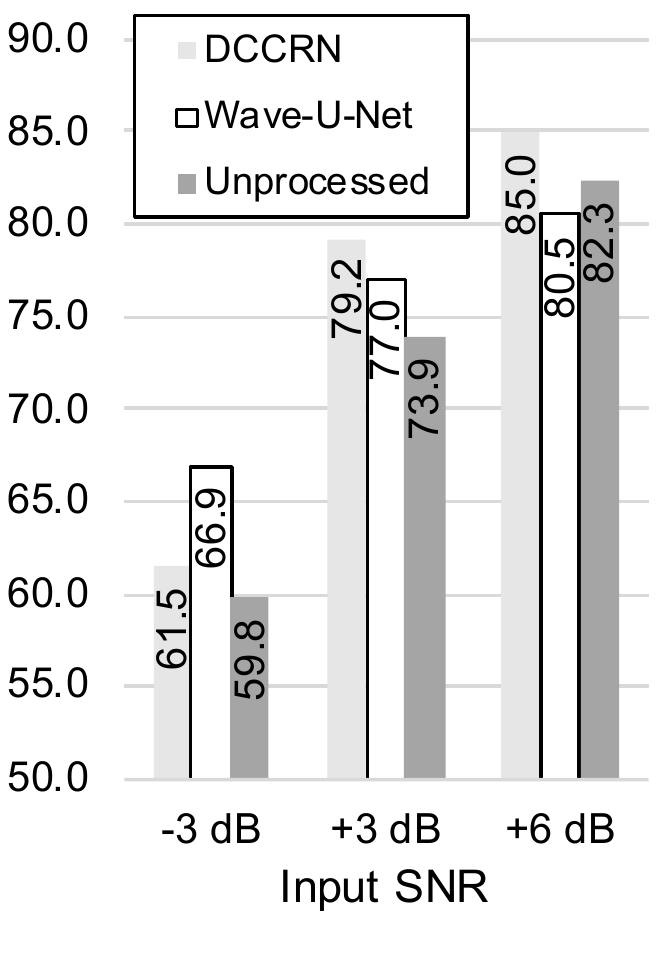}}\hspace{0.0in}
\subfigure[PESQ]{\includegraphics[height=1.75in]{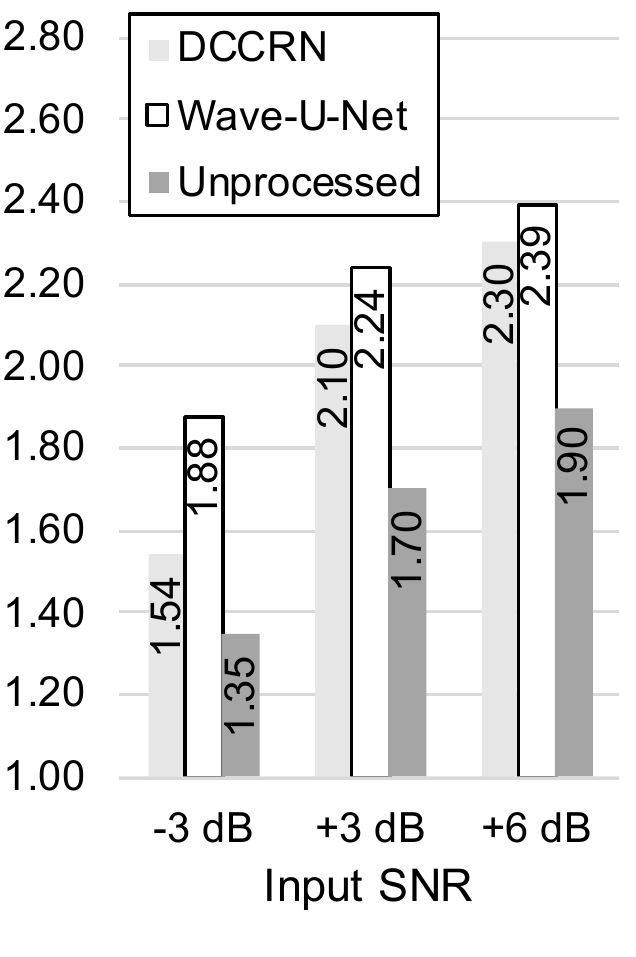}}\hspace{0.0in}
\subfigure[Complexity]{\includegraphics[height=1.75in]{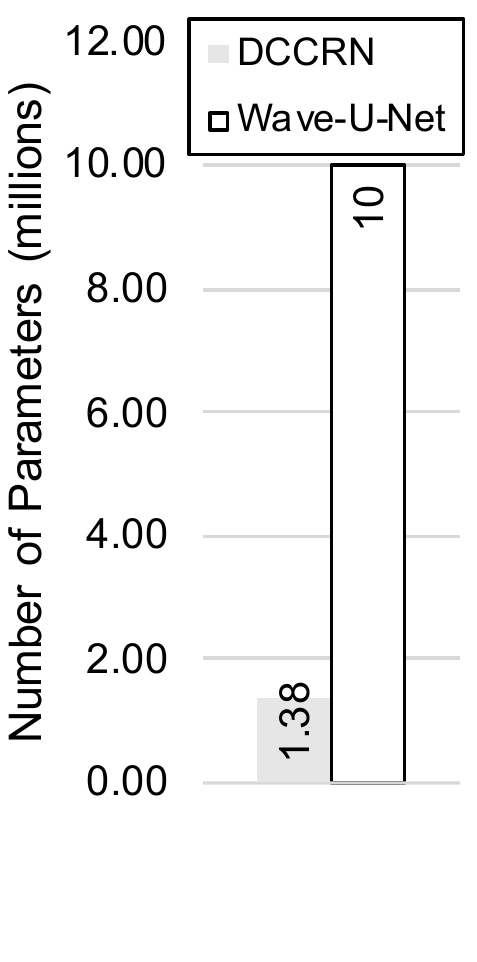}}
\vspace{-0.1in}\caption{Comparisons in terms of STOI, PESQ, model complexity (in million) on untrained speakers and noises against the Wave-U-Net model reported in \cite{liao2019incorporating}}
\label{fig:scaled-up}
\end{figure}

To evaluate model performance in an open condition with unseen speakers and noise sources, we scale up the experimental setting. The training dataset is constructed from 3696 utterances from TIMIT training set. Each utterance is mixed with 100 noise types from \cite{hu2004100} at 6 different SNR levels (20dB, 15dB, 10dB,
5dB, 0dB, and -5dB), which yields 40-hour training data of $\sim$135GB. 100 unseen utterances are randomly selected from TIMIT test set, with each mixed with three untrained noises (Buccaneer1, Destroyer engine, and HF channel from the NOISEX-92 corpus \cite{varga1993assessment}). Note that the performance of Wave-U-Net \cite{stoller2018wave} was reported in terms of STOI and PESQ in the narrowband mode \cite{liao2019incorporating}.

We evaluate models in terms of STOI and PESQ improvements (Fig. \ref{fig:scaled-up} (a) and (b)). Wave-U-Net achieves better speech quality improvement. In terms of speech intelligibility, DCCRN gives higher STOI scores at +3dB and +6dB cases. Note that Wave-U-Net contains about 7.25 times more parameters. Some options to further promote the performance of DCCRN are to expand the size of the receptive field or to incorporate extra phonetic content \cite{liao2019incorporating}, although one of the main focuses of DCCRN is to achieve an affordable model complexity for end-to-end speech enhancement.

\section{Conclusion}
\label{sec:conclusion}
The paper introduces DCCRN, a hybrid residual network, to aggregate temporal context in dual levels for efficient end-to-end speech enhancement. DCCRN firstly suppresses the noise in time domain with dilated DenseNet, followed by a GRU component to further leverage the temporal context in a many-to-one manner. To tune the model with heterogeneity, we present a component-wise training scheme followed by finetuning. Experiments showed that our method consistently outperforms other baseline models in various metrics. It generalizes very well to untrained speakers, and gives reasonable performance on untrained noises with only 1.38 million parameters.



\bibliographystyle{IEEEbib}
\bibliography{refs19}
%
%
%
%
%
%
%
%
%

\end{sloppy}
\end{document}